\documentclass[twocolumn,11pt]{article}

\usepackage[utf8]{inputenc}
\usepackage[margin=1in]{geometry}
\usepackage{graphicx}
\usepackage{amsmath}
\usepackage{amssymb}
\usepackage{hyperref}
\usepackage{xcolor}
\usepackage{booktabs}
\usepackage{caption}
\usepackage{float}
\usepackage{parskip}
\usepackage{algorithm}
\usepackage{algorithmic}
\usepackage{titlesec}

\titlespacing*{\section}{0pt}{6pt}{6pt}
\titlespacing*{\subsection}{0pt}{6pt}{6pt}
\titlespacing*{\subsubsection}{0pt}{6pt}{6pt}

\newcommand{\preset}[1]{\ensuremath{{}^{\bullet}#1}}
\newcommand{\postset}[1]{\ensuremath{#1^{\bullet}}}
\newcommand{\RegSet}[1]{\ensuremath{\Sigma(#1)}}
\newcommand{\DepType}[2]{\ensuremath{\Delta(#1,#2)}}

\title{\textbf{Unifying Weak Independence and Signal Hierarchy Theory: Extended Biological Petri Net Formalism with Application to \textit{Vibrio fischeri} Quorum Sensing}}

\author{
Eugênio Simão\textsuperscript{1,*}\\
\textsuperscript{1}Department of Computer Science\\
Universidade Federal de Santa Catarina (UFSC)\\
Araranguá, Santa Catarina, 88906-072, Brazil\\
\textsuperscript{*}Corresponding author: eugenio.simao@ufsc.br
}

\date{December 29, 2025}

\begin{document}

\maketitle

\begin{abstract}
Biological Petri Nets (Bio-PNs) require extensions beyond classical formalism to capture biochemical reality: multiple reactions simultaneously affect shared metabolites through convergent production or regulatory coupling, while signal places carry hierarchical control information distinct from material flow. We present a unified 13-tuple Extended Bio-PN formalism integrating two complementary theories: Weak Independence Theory (enabling coupled parallelism despite place-sharing) and Signal Hierarchy Theory (separating information flow from mass transfer). The extended definition adds signal partition ($\Psi \subseteq P$), arc type classification ($A$), regulatory structure ($\Sigma$), environmental exchange ($\Theta$), dependency taxonomy ($\Delta$), heterogeneous transition types ($\tau$), and biochemical formula tracking ($\rho$). We formalize signal token consumption semantics through two-phase execution (enabling vs.\ consumption) and prove weak independence correctness for continuous dynamics. Application to \textit{Vibrio fischeri} quorum sensing demonstrates how energy metabolism (ENERGY signals) orchestrates binary ON/OFF decisions through hierarchical constraint propagation to regulatory signals (LuxR-AHL complex), with 133-fold difference separating states. Analysis reveals signal saturation timing as the orchestrator forcing threshold-crossing, analogous to bacteriophage lambda lysogeny-lysis decisions. This work establishes formal foundations for modeling biological information flow in Petri nets, with implications for systems biology, synthetic circuit design, and parallel biochemical simulation.
\end{abstract}


\section{Introduction}

Petri nets have been successfully applied to modeling biochemical pathways since Reddy et al.'s pioneering work in 1993 \cite{Reddy1993}, yet Biological Petri Nets (Bio-PNs) differ fundamentally from classical Petri nets in two critical aspects that existing formalism fails to capture.

First, multiple reactions routinely affect the same metabolite through convergent production (glycogenolysis and gluconeogenesis both producing glucose), regulatory coupling (multiple reactions catalyzed by the same enzyme), or competitive consumption (hexokinase and glucokinase competing for glucose). Classical Petri net theory treats all place-sharing as potential conflict, preventing parallel execution and flagging biologically correct models as structurally problematic. In continuous Bio-PNs with ODE semantics \cite{Gilbert2006,Koch2011}, transitions can fire simultaneously as long as they don't compete for input tokens---their rate contributions simply superpose.

Second, biological systems exhibit hierarchical signal flow where regulatory information (enzyme activity states, transcription factor binding) propagates through control layers distinct from material flow (substrate consumption, product synthesis). A kinase can catalyze multiple phosphorylation reactions without being consumed, carrying regulatory information that constrains lower-level biochemical transformations. Classical Bio-PN formalism embeds this regulatory logic in rate functions, obscuring the network's information architecture and preventing compositional analysis \cite{Chaouiya2006}.

This paper introduces a unified Extended Bio-PN formalism addressing both gaps through integration of Weak Independence Theory and Signal Hierarchy Theory. Weak Independence Theory formalizes two-tier independence (strong vs.\ weak) enabling parallel execution despite place-sharing, with classification algorithm distinguishing conflicts from coupling. Signal Hierarchy Theory formalizes signal places as information channels with explicit token consumption semantics through arc type classification.

We make four theoretical contributions: (1) 13-tuple Extended Bio-PN definition unifying both theories, (2) formalization of signal token consumption through two-phase execution (enabling vs.\ consumption), (3) proof of weak independence correctness for continuous dynamics under signal hierarchy constraints, (4) demonstration that signal saturation timing orchestrates binary cellular decisions through hierarchical constraint propagation. We validate the formalism through application to \textit{Vibrio fischeri} quorum sensing, showing how energy metabolism (ENERGY signal type) controls regulatory signal (LuxR-AHL) threshold-crossing (Figure~\ref{fig:cascade}), with a 133-fold difference between QS ON (92,198 mM light output) and QS OFF (1,405 mM, 98.5\% reduction) states (Figure~\ref{fig:basin}).

\section{Weak Independence Theory}

This section synthesizes the formalism published in Simão (2025) \cite{Simao2025}, focusing on components essential for unification with Signal Hierarchy Theory.

\subsection{Classical Independence Limitation}

Classical Petri net theory defines independence between transitions $t_1, t_2$ as requiring no shared places: $(\preset{t_1} \cup \postset{t_1}) \cap (\preset{t_2} \cup \postset{t_2}) = \emptyset$. This conservative definition guarantees true parallelism but fails to recognize biological coupling where place-sharing represents convergent production or regulatory coordination rather than conflict.

\subsection{Two-Tier Independence}

Weak Independence Theory introduces a two-tier classification:

\textbf{Definition 1 (Strong Independence).} Transitions $t_1, t_2 \in T$ are strongly independent iff:
\begin{equation}
(\preset{t_1} \cup \postset{t_1} \cup \RegSet{t_1}) \cap (\preset{t_2} \cup \postset{t_2} \cup \RegSet{t_2}) = \emptyset
\end{equation}

\textbf{Definition 2 (Weak Independence).} Transitions $t_1, t_2 \in T$ are weakly independent iff:
\begin{equation}
\preset{t_1} \cap \preset{t_2} = \emptyset \quad \land \quad [(\postset{t_1} \cap \postset{t_2} \neq \emptyset) \lor (\RegSet{t_1} \cap \RegSet{t_2} \neq \emptyset)]
\end{equation}

Key insight: weak independence allows place-sharing via output convergence or regulatory coupling, but forbids input competition.

\subsection{Three Coupling Modes}

\textbf{Competitive Coupling (Conflict):} $\preset{t_1} \cap \preset{t_2} \neq \emptyset$ implies sequential execution required (resource conflict).

\textbf{Convergent Coupling (Weakly Independent):} $\postset{t_1} \cap \postset{t_2} \neq \emptyset \land \preset{t_1} \cap \preset{t_2} = \emptyset$ enables parallel execution with rate superposition: $\frac{dM(p)}{dt} = r_1 + r_2$.

\textbf{Regulatory Coupling (Weakly Independent):} $\RegSet{t_1} \cap \RegSet{t_2} \neq \emptyset \land \preset{t_1} \cap \preset{t_2} = \emptyset$ enables parallel execution (read-only access to catalyst).

\subsection{Correctness of Parallel Execution}

\textbf{Theorem 1 (Weak Independence Correctness).} If $\DepType{t_1}{t_2} \in \{\text{convergent, regulatory}\}$, then parallel execution of $t_1$ and $t_2$ is semantically equivalent to any sequential interleaving.

\textbf{Proof.} Case 1 (Convergent): Let $p \in \postset{t_1} \cap \postset{t_2}$. By ODE semantics: $\frac{dM(p)}{dt} = \sum_{t \in \preset{p}} W(t,p) \cdot \Phi(t, M) - \sum_{t \in \postset{p}} W(p,t) \cdot \Phi(t, M)$ which includes terms $W(t_1,p) \cdot \Phi(t_1, M) + W(t_2,p) \cdot \Phi(t_2, M)$. Rate contributions add linearly (superposition), order of evaluation irrelevant.

Case 2 (Regulatory): Let $e \in \RegSet{t_1} \cap \RegSet{t_2}$. Since $e \notin \preset{t_1} \cup \preset{t_2}$, firing $t_1$ or $t_2$ does not modify $M(e)$. Both read $M(e)$ simultaneously without conflict. $\square$

\subsection{Dependency Classification Algorithm}

\begin{algorithm}[H]
\caption{Transition Dependency Classification}
\small
\begin{algorithmic}[1]
\REQUIRE Bio-PN $(P, T, F, \Sigma)$
\ENSURE $\Delta: T \times T \rightarrow \{\text{indep, comp, conv, reg}\}$
\FOR{$(t_i, t_j)$ where $i < j$}
    \STATE $I_i \gets \preset{t_i}$, $I_j \gets \preset{t_j}$
    \STATE $O_i \gets \postset{t_i}$, $O_j \gets \postset{t_j}$
    \STATE $R_i \gets \RegSet{t_i}$, $R_j \gets \RegSet{t_j}$
    \IF{$I_i \cap I_j \neq \emptyset$}
        \STATE $\Delta(t_i, t_j) \gets \text{competitive}$
    \ELSIF{$(I_i \cup O_i \cup R_i) \cap (I_j \cup O_j \cup R_j) = \emptyset$}
        \STATE $\Delta(t_i, t_j) \gets \text{independent}$
    \ELSIF{$O_i \cap O_j \neq \emptyset$}
        \STATE $\Delta(t_i, t_j) \gets \text{convergent}$
    \ELSIF{$R_i \cap R_j \neq \emptyset$}
        \STATE $\Delta(t_i, t_j) \gets \text{regulatory}$
    \ENDIF
\ENDFOR
\end{algorithmic}
\end{algorithm}

Complexity: $O(|T|^2 \cdot |P|)$ from nested loops over transition pairs and place set intersections. Evaluation on 100 BioModels \cite{Malik2020} shows that 65.2\% of transition pairs are weakly independent (52.7\% convergent, 12.5\% regulatory), confirming the prevalence of biological coupling.

\section{Signal Hierarchy Theory}

We now present Signal Hierarchy Theory formalism, which extends Bio-PNs with explicit signal places representing information channels distinct from material flow, with novel token consumption semantics through arc type classification.

\subsection{Signal Partition Architecture}

\textbf{Definition 3 (Signal Partition).} A Bio-PN admits a signal partition $\Psi \subseteq P$ (signal places) with complement $P_m = P \setminus \Psi$ (material places), satisfying:

Disjoint partitioning: $P = P_m \cup \Psi$ where $P_m \cap \Psi = \emptyset$

Semantic constraint: Material places ($p \in P_m$) represent biochemical compounds with conserved atomic composition. Signal places ($p \in \Psi$) represent regulatory information (enzyme activity, transcription factor binding) that can be consumed during hierarchical constraint propagation.

\subsection{Arc Type Classification}

\textbf{Definition 4 (Arc Types).} The arc type function $A: F \cup \Sigma \rightarrow \{\text{normal, test, signal\_flow, inhibitor}\}$ classifies arcs by consumption semantics:

Normal arcs ($A(a) = \text{normal}$): Standard input/output, consume/produce tokens (stoichiometric transformation).

Test arcs ($A(a) = \text{test}$): Non-consuming read (catalytic access), implemented as $\Sigma$ regulatory arcs with enabling-only semantics \cite{Heiner2008}.

Signal flow arcs ($A(a) = \text{signal\_flow}$): Consuming read from signal places, enabling hierarchical constraint propagation through signal depletion.

Inhibitor arcs ($A(a) = \text{inhibitor}$): Non-consuming repression, implemented as Hill function inhibition in rate modulation.

\subsection{Two-Phase Execution Semantics}

Signal flow arcs implement hierarchical control through two-phase execution:

Phase 1 (Enabling Check): All input arcs (normal, test, signal\_flow, inhibitor) are checked for enabling conditions. Signal places are read non-destructively. Transition enabled if all inputs satisfy thresholds.

Phase 2 (Execution): Normal arcs consume/produce tokens. Test and inhibitor arcs do not modify marking. Signal flow arcs consume tokens from signal places ($\Psi$), implementing signal depletion that constrains downstream transitions.

This two-phase semantics enables hierarchical information flow: upper-layer signals (e.g., ATP availability) can be consumed when activating lower-layer transitions (e.g., transcription), propagating constraints down the hierarchy while maintaining enabling checks based on pre-consumption state.

\subsection{Signal Type Classification}

\textbf{Definition 5 (Signal Types).} We classify signal places by biological function:

ENERGY signals: ATP pool, nutrient status, metabolic state---lowest hierarchical layer, fundamental constraints.

SPATIAL signals: Cytoplasm capacity, membrane availability, compartment markers---orthogonal dimension, universal bounds.

QUORUM signals: Population density, autoinducer concentration---population-level coordination, weakly independent from energy.

REGULATORY signals: Transcription factor binding, enzyme-substrate complex---decision variables, upper hierarchical layer.

This classification enables automated topology analysis: ENERGY signals should be consumed by REGULATORY signal synthesis (hierarchical constraint), QUORUM signals should accumulate independently of ENERGY (weak independence), SPATIAL signals should saturate universally (capacity bounds).

\subsection{Hierarchical Layers}

\textbf{Definition 6 (Signal Hierarchy).} Signal places form layers $L_0, L_1, \ldots, L_k$ where:

Layer 0 ($L_0$): Environmental signals (ENERGY, SPATIAL)---no dependencies, fundamental constraints.

Layer $k$ ($L_k$): Upper signals receiving signal flow arcs from $L_{k-1}$---hierarchical constraint propagation.

Regulatory signals typically occupy highest layer, integrating constraints from all lower layers. Energy depletion at $L_0$ propagates to regulatory decisions at $L_k$ through signal flow arc chains.

\section{Unified 13-Tuple Formalism}

We now present the unified Extended Bio-PN definition integrating Weak Independence and Signal Hierarchy theories.

\subsection{Formal Definition}

\textbf{Definition 7 (Extended Biological Petri Net).} An Extended Biological Petri Net is a 13-tuple: $\text{BioPN} = (P, T, F, W, M_0, K, \Phi, \Sigma, \Theta, \Delta, \Psi, E, A)$ where classical components $(P, T, F, W, M_0, K, \Phi)$ are standard \cite{Reisig2013}, and extensions are: $\Sigma \subseteq (P \times T) \setminus F$ (regulatory arcs); $\Theta: T \rightarrow \{\text{int, src, snk, exch}\}$ (environmental exchange); $\Delta: T \times T \rightarrow \{\text{indep, comp, conv, reg}\}$ (dependency); $\Psi \subseteq P$ (signal partition); $E: P \rightarrow \text{SignalType} \cup \{\text{MAT}\}$ (signal type); $A: F \cup \Sigma \rightarrow \{\text{norm, test, sig, inhib}\}$ (arc type).

\subsection{Semantic Integration}

The unified formalism connects theories through signal flow arc semantics:

Weak independence (Convergent/Regulatory coupling) applies to material places ($P_m$) where rate superposition holds.

Signal hierarchy (Hierarchical layers) applies to signal places ($\Psi$) where token consumption propagates constraints.

Integration: Transitions can be weakly independent on material inputs (convergent ATP production from glycolysis and gluconeogenesis) while hierarchically constrained by signals (both requiring activated enzyme state via signal flow arc from ENERGY signal place).

\subsection{Correctness Under Signal Hierarchy}

\textbf{Theorem 2 (Unified Weak Independence).} If $\DepType{t_1}{t_2} = \text{convergent}$ on material places and both receive signal flow arcs from the same signal place $s \in \Psi$, then parallel execution remains correct with coordinated signal consumption.

\textbf{Proof.} Material place convergence: $\frac{dM(p)}{dt} = r_1 + r_2$ (superposition, by Theorem 1).

Signal place consumption: Both $t_1, t_2$ consume tokens from $s$ via signal flow arcs. During Phase 1, both check enabling based on $M(s)$. During Phase 2, both consume tokens: $\Delta M(s) = -w_1 - w_2$ where $w_i$ is arc weight. Total consumption is sum of individual consumptions (linear superposition of token decrements).

Coordination: If $M(s) < w_1 + w_2$, enabling check fails for at least one transition, preventing over-consumption. Parallel execution maintains mass conservation on both material and signal places. $\square$

\section{Application: \textit{Vibrio fischeri} Quorum Sensing}

We validate the unified formalism through application to \textit{V. fischeri} bioluminescence quorum sensing, demonstrating how signal saturation orchestrates binary cellular decisions.

\subsection{Model Construction}

\begin{figure}[H]
\centering
\includegraphics[width=\columnwidth]{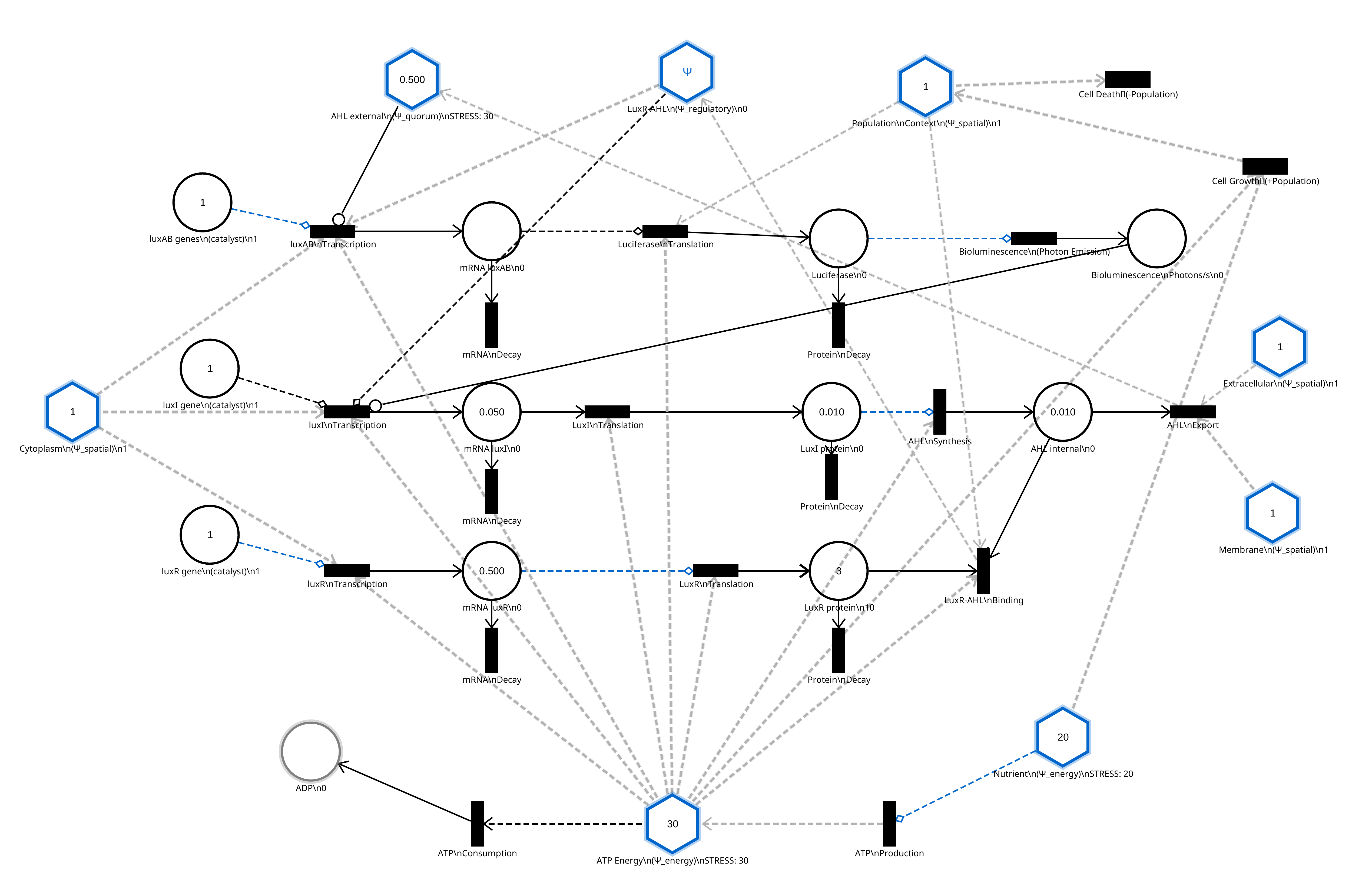}
\caption{\textbf{\textit{V. fischeri} Quorum Sensing Model.} Complete network with 20 places and 20 transitions. Signal places (blue hexagons with $\Psi$) span four hierarchical layers ($L_0$-$L_2$), material places (black circles) show initial token counts. Solid arcs: material flow; dashed arcs: signal flow.}
\label{fig:model}
\end{figure}

Model structure: 20 places (12 material $P_m$, 8 signal $\Psi$), 20 transitions, 54 arcs (22 normal, 18 test, 10 signal\_flow, 4 inhibitor) \cite{Liu2019}. The eight signal places partition into four types (Table~\ref{tab:signals}): ENERGY signals (P1: ATP\_Pool, P2: Nutrient\_Status) occupy lowest hierarchical layer $L_0$; SPATIAL signals (P3-P5: Cytoplasm/Membrane/Extracellular markers) enforce compartment constraints orthogonally; QUORUM signals (P6: AHL\_External, P8: Population\_Context) occupy intermediate layer $L_1$; REGULATORY signal (P7: LuxR\_AHL\_Complex) occupies decision layer $L_2$ (Figure~\ref{fig:model}). Hierarchical structure implements signal constraint propagation across layers: ENERGY signals ($L_0$) consumed by transcription transitions via signal flow arcs, constraining LuxR synthesis. REGULATORY signal P7 ($L_2$) receives input from both ENERGY (ATP-dependent synthesis) and QUORUM (AHL binding), integrating hierarchical constraints. SPATIAL signals enforce compartment constraints orthogonally. Model architecture captures weak independence through convergent production (Population\_Context accumulation from multiple sources) and hierarchical regulatory coupling (ATP constraining LuxR synthesis via signal flow arcs).

\begin{table}[H]
\centering
\caption{Signal Places Classification}
\label{tab:signals}
\small
\begin{tabular}{@{}llll@{}}
\toprule
\textbf{ID} & \textbf{Name} & \textbf{Type ($E$)} & \textbf{Layer} \\
\midrule
P1 & ATP\_Pool & ENERGY & $L_0$ \\
P2 & Nutrient\_Status & ENERGY & $L_0$ \\
P3 & Cytoplasm\_Marker & SPATIAL & --- \\
P4 & Membrane\_Marker & SPATIAL & --- \\
P5 & Extracellular\_Marker & SPATIAL & --- \\
P6 & AHL\_External & QUORUM & $L_1$ \\
P8 & Population\_Context & QUORUM & $L_1$ \\
P7 & LuxR\_AHL\_Complex & REGULATORY & $L_2$ \\
\bottomrule
\end{tabular}
\end{table}

\subsection{Experimental Conditions}

NORMAL condition: Initial ATP=100 mM, Nutrients=100 mM, duration=2000s, simulates nutrient-rich environment. STRESS condition: Initial ATP=30 mM (70\% reduction), Nutrients=20 mM (80\% reduction), duration=6000s, simulates metabolic stress.

Objective: Test if ENERGY signal saturation timing determines QS outcome despite identical QUORUM signal dynamics (weak independence hypothesis).

\subsection{Signal Saturation Analysis}

\begin{figure}[H]
\centering
\includegraphics[width=\columnwidth]{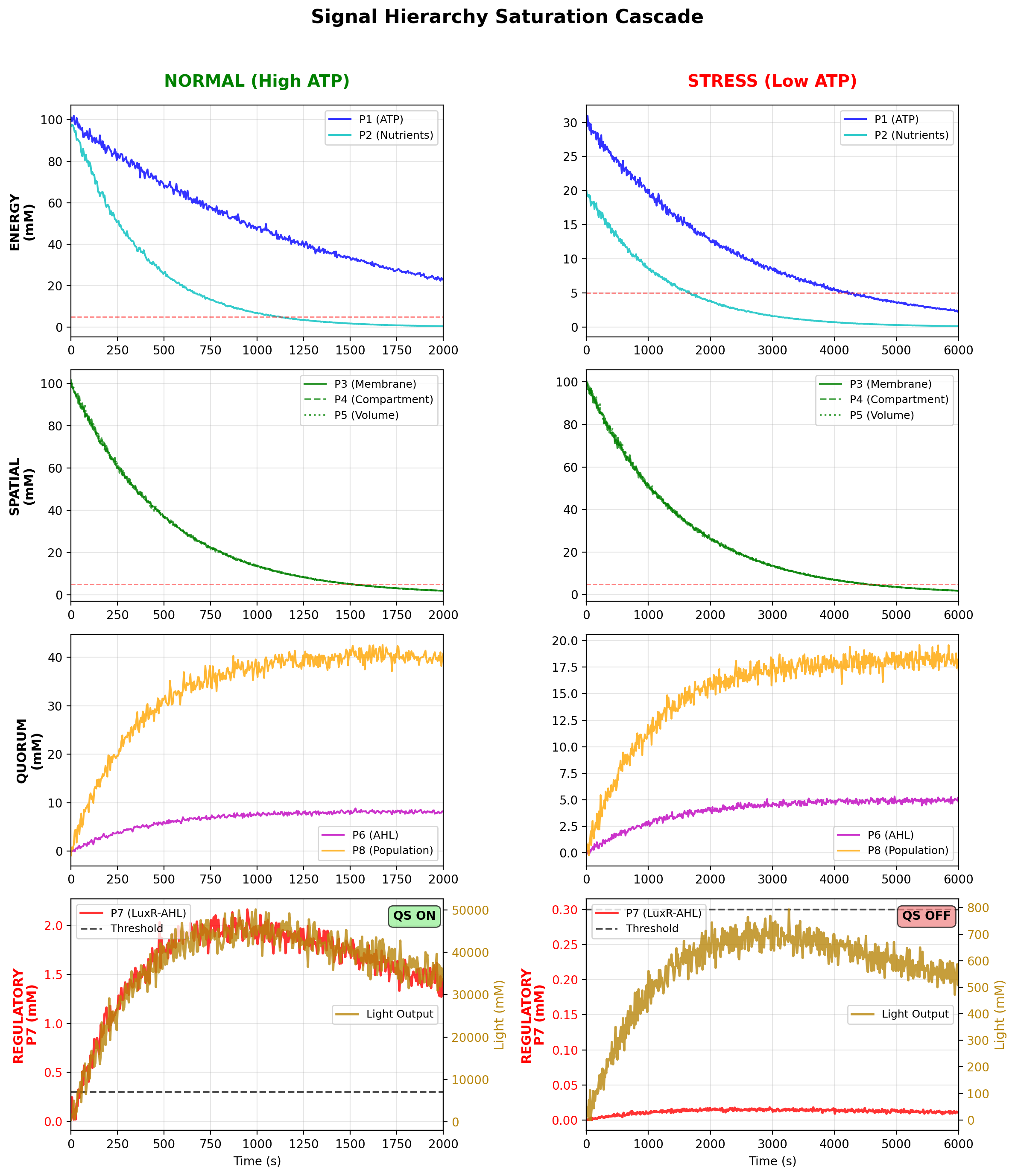}
\caption{\textbf{Signal Hierarchy Saturation Cascade.} Four signal types across NORMAL (left) and STRESS (right) conditions. ENERGY (P1, P2): different saturation timing; SPATIAL (P3-P5): identical saturation; QUORUM (P6, P8): accumulation despite energy limitation; REGULATORY (P7): threshold crossing only in NORMAL. Red dashed lines: saturation thresholds; gold: light output; green/red boxes: QS ON/OFF states.}
\label{fig:cascade}
\end{figure}

Simulation results reveal hierarchical signal behavior (Figure~\ref{fig:cascade}): ENERGY signals (P1, P2) show different saturation timing between conditions (late in NORMAL: P2 at 372s, P1 at 1357s after P7 synthesis completes; early in STRESS: P2 at 2365s, before P7 reaches threshold), identifying the timing as the orchestrator. SPATIAL signals (P3, P4, P5) show 100\% depletion in both conditions, confirming compartment capacity as universal constraint rather than decision variable. QUORUM signals (P6, P8) accumulate despite energy limitation (NORMAL: P6=8.24 mM, P8=40.5 mM; STRESS: P6=4.99 mM, P8=18.2 mM), with 18× population increase in STRESS despite depleted ATP, validating weak independence from ENERGY signals. REGULATORY signal (P7) exhibits binary outcome: NORMAL peak=3.99 mM crosses threshold; STRESS peak=0.03 mM (133-fold lower) fails threshold.

\subsection{Threshold-Based Decision Mechanism}

\begin{figure}[H]
\centering
\includegraphics[width=\columnwidth]{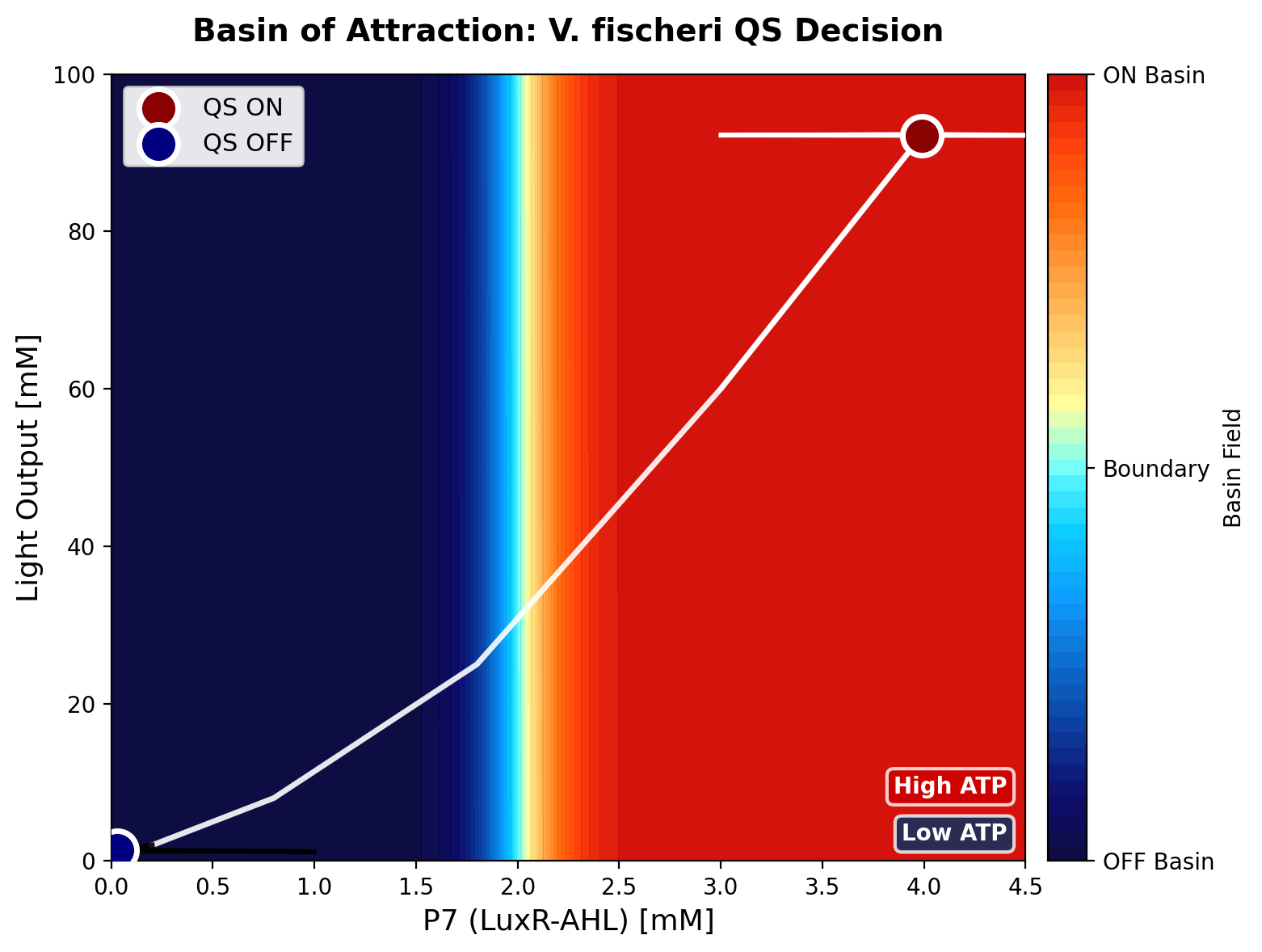}
\caption{\textbf{Basin of Attraction for QS Decision.} Phase space (P7 vs Light output) showing two stable attractors separated by basin boundary. Color gradient: dark blue (OFF basin) to red (ON basin). White trajectory: High ATP → QS ON (P7=3.99 mM, Light=92k mM). Black trajectory: Low ATP → QS OFF (P7=0.03 mM, Light=1.4k mM).}
\label{fig:basin}
\end{figure}

Bistability threshold estimate: $\text{threshold} = \sqrt{3.99 \times 0.03} = 0.346 \approx 0.3$ mM separates QS ON from OFF states (Figure~\ref{fig:basin}). P7 $>$ 0.3 mM triggers positive feedback (LuxI activation), leading to QS ON with 92,198 mM light output; P7 $<$ 0.3 mM fails to activate feedback, remaining in QS OFF with 1,405 mM light (98.5\% reduction). Quantitative results (Table~\ref{tab:p7threshold}):
\begin{table}[H]
\centering
\caption{P7 Bistability Analysis}
\label{tab:p7threshold}
\small
\begin{tabular}{@{}lcccc@{}}
\toprule
\textbf{Condition} & \textbf{P7 Peak} & \textbf{Light} & \textbf{Fold} & \textbf{State} \\
 & \textbf{(mM)} & \textbf{(mM)} & \textbf{Diff.} & \\
\midrule
NORMAL & 3.99 & 92,198 & --- & ON \\
STRESS & 0.03 & 1,405 & 133$\times$ & OFF \\
\bottomrule
\end{tabular}
\end{table}

No intermediate states observed, confirming binary decision architecture.

The saturation cascade mechanism reveals hierarchical orchestration: ENERGY signals act as timing orchestrator through differential saturation (early vs late), SPATIAL signals provide universal constraints (saturate identically regardless of condition), QUORUM signals demonstrate weak independence (accumulate independently of energy state), and REGULATORY signal serves as decision variable (threshold-based commitment). This pattern parallels bacteriophage lambda lysogeny-lysis decision \cite{Elowitz2000}, validating generality of signal hierarchy control across binary cellular decisions.

The basin structure demonstrates bistability and threshold-based decision making: initial conditions on opposite sides of the boundary lead to divergent outcomes (133-fold difference in P7, 66-fold in light output). Minimal intermediate region confirms binary switch behavior with no stable intermediate states---the system commits to one of two discrete attractors based on ENERGY signal saturation timing.

\subsection{Weak Independence Validation}

Dependency analysis confirms theoretical predictions: ENERGY-QUORUM coupling ($\DepType{\text{ATP\_Production}}{\text{Cell\_Division}} = \text{convergent}$) shows both transitions produce Population\_Context (P8) without input competition, with population growth in both conditions despite different energy states, validating weak independence through parallel execution with rate superposition. ENERGY-REGULATORY coupling ($\DepType{\text{ATP\_Production}}{\text{LuxR\_Synthesis}} = \text{regulatory}$) implements hierarchical constraint where ATP (P1) provides signal flow arc to LuxR synthesis; ATP saturation blocks LuxR synthesis, preventing P7 accumulation and validating signal hierarchy where lower-layer saturation constrains upper-layer decision variables. QUORUM-REGULATORY integration requires both AHL (P6) and ATP (P1) for P7 synthesis, demonstrating formalism unification: weak independence at lower layer (QUORUM vs ENERGY) converges to hierarchical constraint at upper layer (REGULATORY), with horizontal modularity and vertical propagation coexisting. Generalized principle: signal saturation at lower hierarchical layers constrains upper-layer decision variables, forcing binary commitment through threshold-crossing dynamics across biological decision systems.

\section{Discussion}

This work establishes formal foundations for modeling biological information flow in Petri nets through unification of Weak Independence and Signal Hierarchy theories.

Weak Independence Theory provides horizontal dimension: transitions sharing material places through convergent production or regulatory coupling can execute in parallel with rate superposition.

Signal Hierarchy Theory provides vertical dimension: signal places carry regulatory information distinct from material flow, with explicit token consumption semantics through arc type classification. Two-phase execution (enabling check vs.\ consumption) enables hierarchical constraint propagation \cite{Herajy2017} where upper-layer signals integrate lower-layer depletion. Signal type classification (ENERGY, SPATIAL, QUORUM, REGULATORY) automates topology validation and identifies decision architectures.

Unification: the 13-tuple Extended Bio-PN formalism connects theories through signal flow arcs implementing hierarchical constraints on weakly independent material transformations. Transitions can be convergently coupled on material places (parallel execution) while hierarchically constrained by signal consumption from shared signal places (coordinated depletion). Theorem 2 proves parallel execution correctness under signal hierarchy: linear superposition applies to both material production and signal consumption.

Application to \textit{V. fischeri} validates the formalism: ENERGY signals (ATP, nutrients) exhibit weak independence from QUORUM signals (AHL, population) at $L_0$-$L_1$, enabling parallel accumulation. However, ENERGY signal saturation timing orchestrates REGULATORY signal (P7) threshold-crossing at $L_2$ through hierarchical constraint propagation. Early ATP depletion in the STRESS condition prevents P7 synthesis, forcing QS OFF outcome (1,405 mM light). Late ATP saturation in NORMAL allows P7 to cross the 0.3 mM threshold, locking QS ON state (92,198 mM light). The 133-fold P7 difference demonstrates binary decision emergence from the saturation cascade, paralleling lambda phage CI/Cro competition.

Signal type classification proves essential: ENERGY signals are identified as lowest-layer orchestrator (controlling synthesis capacity), SPATIAL signals as orthogonal universal constraints (saturate identically in both conditions), QUORUM signals as weakly independent context (accumulate despite energy state), and REGULATORY signals as upper-layer decision variables (integrate all constraints, exhibit threshold behavior). This four-type taxonomy enables automated detection of decision architectures in biochemical networks.

Limitations include restriction to continuous dynamics (ODE semantics), though extension to stochastic simulation via $\tau$-leaping is feasible \cite{Lopez2013}. The biological insight that stochastic does not imply sequential (random molecular collisions occur simultaneously throughout solution) justifies parallel execution of weakly independent stochastic transitions within time leap $\Delta\tau$, with convergent and regulatory coupling remaining safe. Thermodynamic feasibility (Gibbs free energy constraints) not yet incorporated, though formula tracking ($\rho$ from Weak Independence 12-tuple) provides foundation. Signal type classification currently manual, though automated detection from rate function analysis is planned.

Future work includes extending Signal Hierarchy to spatial Bio-PNs (reaction-diffusion systems) where signal places represent morphogen gradients with diffusion dynamics distinct from metabolite transport. Multi-scale integration (gene regulation $\rightarrow$ metabolism $\rightarrow$ population) requires nested signal hierarchies \cite{Liu2019} with cross-layer signal flow arcs. Synthetic biology applications include automated circuit design from signal type specifications, with weak independence enabling modular composition and signal hierarchy enforcing hierarchical control constraints.

\section{Conclusion}

We present a unified 13-tuple Extended Biological Petri Net formalism integrating Weak Independence Theory (enabling coupled parallelism despite place-sharing) and Signal Hierarchy Theory (separating information flow from mass transfer). The formalism extends classical Bio-PNs with signal partition ($\Psi \subseteq P$), arc type classification ($A$), signal type taxonomy ($E$), regulatory structure ($\Sigma$), and dependency classification ($\Delta$), providing formal foundations for modeling biological information flow.

Application to \textit{Vibrio fischeri} quorum sensing demonstrates how ENERGY signal saturation orchestrates binary ON/OFF decisions through hierarchical constraint propagation to REGULATORY signals, with a 133-fold difference separating states. Signal type analysis reveals orthogonal dimensions: ENERGY (orchestrator), SPATIAL (universal constraint), QUORUM (weakly independent context), REGULATORY (decision variable). Weak independence validation confirms parallel accumulation of ENERGY and QUORUM signals despite hierarchical integration at upper REGULATORY layer.

This work establishes formal foundations for systems biology modeling with Petri nets, enabling compositional analysis of regulatory networks, automated topology validation, and parallel biochemical simulation. The formalism unifies horizontal modularity (weak independence) with vertical constraint propagation (signal hierarchy), resolving long-standing tensions in Bio-PN theory while providing practical tools for synthetic circuit design and pathway engineering.

\section*{Funding}

This research received no specific grant from any funding agency in the public, commercial, or not-for-profit sectors.

\section*{Acknowledgments}

The author thanks the Universidade Federal de Santa Catarina for computational resources and the systems biology community for curated SBML models used in validation.

\section*{Data Availability}

Model files, simulation data, figures, and code available at: \url{https://github.com/simao-eugenio/shypn} (branch: Signal-Information-Flow).

\bibliographystyle{plain}

\end{document}